\documentstyle[12pt]{article}

\begin{document}
\begin{titlepage}
\setcounter{page}{1}
\title{Hamiltonian quantization \\ of General Relativity \\ with the
change of signature}
\author{J\'er\^ome Martin\thanks{boursier du Minist\`ere de la Recherche et
de l'Espace.}\\
\mbox{\small Laboratoire de Gravitation et Cosmologie Relativistes} \\
\mbox{\small Universit\'e Pierre et Marie Curie, CNRS/URA 769} \\
\mbox{\small Tour 22/12, Boite Courrier 142} \\
\mbox{\small 4, place Jussieu 75252 Paris Cedex 05, FRANCE}}
\maketitle

\begin{abstract}
We show in this article how the usual hamiltonian formalism of
General Relativity should be modified in order to allow the inclusion of
the Euclidean classical solutions of Einstein's equations. We study the
effect that the dynamical change of signature has on the superspace and we
prove that it induces a passage of the signature of the supermetric
from ($-+++++$) to ($+-----$). Next, all these features are more particularly
studied on the example of minisuperspaces. Finally, we consider the problem
of quantization of the Euclidean solutions. The consequences of different
choices of boundary conditions are examined.
\end{abstract}
\end{titlepage}

\section{Introduction}

One of the most important results in quantum cosmology has been the suggestion
that the very early Universe might be described by an Euclidean manifold.
This idea arises from the interpretation of the behaviour of wave function
of the Universe which is very similar to the behaviour of the wave function
of a particle in a classically forbidden region. Thus, Euclidean Universes
should be a natural feature of quantum cosmology. However, recent works have
shown that solutions of Einstein's equations with change of signature might
be possible \cite{r1,r2,r3,r4} also as a part of classical theory. The
Hamiltonian formulation of General Relativity (ADM formalism) \cite{r5} is
particularly well adapted to the study of the signature change. Indeed, it
is conceived in the very beginning as a slicing of the four-dimensional
spacetime into three-dimensional spacelike hypersurfaces. The geometry of
these hypersurfaces contains the true degrees of freedom of the theory,
whereas the way how the hypersurfaces are stacked together is essentially
arbitrary and depends on the dynamics.
\par
One of the aims of this article is to show how we can construct a slightly
modified usual Hamiltonian formalism in order to incorporate
classical Euclidean solutions. Then, we study the consequences of the
dynamical change of signature on the metric of the superspace. We prove
that Euclidean solutions correspond to a region in the superspace
where signature of the metric is ($+-----$) instead of ($-+++++$). Finally,
we consider the quantization of this model of gravitation. Since the
classical Euclidean solutions do exist, it seems interesting to find out how
the corresponding wave function behaves and to compare its behaviour with
the behaviour of the wave functions which are interpreted as describing
Euclidean regions of the Universe. Another motivation is that the change
of signature could occur during the Planck epoch \cite{r1}. In that case
the quantum theoretical approach would become necessary.

\section{Hamiltonian formalism}

In this section we show how classical Einstein's solutions with
change of signature can be obtained by means of a slight modification of the
ADM formalism of General Relativity. We consider a four-dimensional manifold
($M$, $g_{\mu \nu}$)\footnote{In this article, Greek indices run from 0 to 3
whereas Latin indices run from 1 to 3} whose structure can be symbolized by:
\begin{equation}
\label{1}
M=M^+\cup \Sigma \cup M^-
\end{equation}
where $M^+$ is a submanifold endowed with an Euclidean metric
(signature $++++$) whereas $M^-$ is endowed with a Lorentzian metric
(signature $-++++$). These two submanifolds are matched together at $\Sigma$,
the surface of change of the signature on which $g_{\mu \nu}$ is degenerate.
An additional difficulty arises while considering the ADM formalism since
the basic idea of the Hamiltonian formulation of General Relativity consists
in slicing along the timelike coordinate the four-dimensional manifold
into three-dimensional spacelike hypersurfaces and in following the evolution
of these hypersurfaces in time. In the Euclidean region, space and time are
completely equivalent and there is no a priori direction which could be
considered as being the coordinate with respect to which the slicing should
be performed. However, the time coordinate of the Lorentzian region induces
in the vicinity of the separating surface $\Sigma$ a privileged direction in
the Euclidean region which will be chosen in order to slice the submanifold
$M^+$. Following the notations adopted by Ellis et al. \cite{r1}, let us
define the symbol $\epsilon$ to be equal to $-1$ on $M^+$ and $+1$
on $M^-$:
\begin{equation}
\label{2}
\left\{\begin{array}{ll}
\epsilon=-1 \ \ \mbox{on $M^+$} \\
\epsilon=+1 \ \ \mbox{on $M^-$}
\end{array}\right.
\end{equation}
The three-dimensional hypersurfaces are endowed with the metric $h_{ij}$.
Then, the metric $g_{\mu \nu}$ can be written as:
\begin{eqnarray}
\label{3}
{\rm d}s^2 &=& g_{\mu \nu}{\rm d}x^{\mu}{\rm d}x^{\nu} \\
\label{4}
           &=& -\epsilon(\sqrt{\epsilon N}{\rm d}t)^2+h_{ij}({\rm d}x^i+N^i
           {\rm d}t)({\rm d}x^j+N^j{\rm d}t)
\end{eqnarray}
where $N(x^{\mu })$ and $N^i(x^{\mu })$ are respectively the usual lapse
and shift functions of the ADM formalism \cite{r5}. The term $\epsilon$
in front of the first term takes into
account the fact that the distance between two neighbour points (nothing else
than the Pythagoras' theorem!) is calculated differently according to the
signature. We note that the proper time elapsed between two points with the
same spacelike coordinates is now given by the formula:
\begin{equation}
\label{4.1}
{\rm d}\tau=\sqrt{\epsilon N}{\rm d}t
\end{equation}
Then, the metric of the manifold $M$ can be written as:
\begin{equation}
\label{5}
{\rm d}s^2=(N^iN_i-N){\rm d}t^2+2N_i{\rm d}x^i{\rm d}t+h_{ij}{\rm d}x^i
{\rm d}x^j
\end{equation}
with $N$ allowed to take on any real values, which leads to the possibility
of including the signature changes. The inverse metric is:
\[\left(\begin{array}{clcr}
\label{6}
-\frac{1}{N} & \ \ \ \frac{N^i}{N} \\
\frac{N^i}{N} & h^{ij}-\frac{N^iN^j}{N}
\end{array} \right) \]
We have to examine now how the extrinsic curvature can be expressed in such a
manifold. Let $n$ be the vector perpendicular to the congruence of
three-dimensional hypersurfaces, $\partial _t$ the vector associated to the
timelike coordinate $t$ and $\partial _i$ three basic (coordinate) vectors
of the spacelike hypersurfaces. The obvious relation holds:
\begin{equation}
\label{7}
\partial _t=\sqrt{\epsilon N}n+N^i\partial _i,
\end{equation}
from which follows that the contravariant and covariant components of the
vector $n$ are:
\begin{eqnarray}
\label{8}
n^{\mu} &=& \frac{1}{\sqrt{\epsilon N}}(1,-N^i) \\
\label{9}
n_{\mu} &=& (-\epsilon \sqrt{\epsilon N},0,0,0)
\end{eqnarray}
To compute the extrinsic curvature, we use the following
coordinate-independent formula:
\begin{equation}
\label{10}
K_{ij}=\frac{1}{2}{\cal L}_n h_{ij}
\end{equation}
where ${\cal L} _n$ is the Lie derivative with respect to the vector field
$n$. In terms of lapse and shift functions, the above relation can be
written as:
\begin{eqnarray}
\label{11}
K_{ij} &=& \epsilon \sqrt{\epsilon N} \Gamma^0_{ij} \\
\label{12}
       &=& -\frac{1}{2\sqrt{\epsilon N}}(^{(3)}\nabla _{(i}N_{j)}-\partial
       _t h_{ij})
\end{eqnarray}
where $^{(3)}\nabla $ is the intrinsic covariant derivative on the
three-dimensional spacelike hypersurfaces. Using the expression (\ref{5})
of the metric, we can now compute the components of the Ricci tensor. They
are given by:
\begin{eqnarray}
\label{19}
R_{00} &=& -\sqrt{\epsilon N}h^{ki}\stackrel{\mbox{\LARGE .}}{K}_{ki}
       +\frac{1}{2}{}^{(3)}\nabla _k
       (\partial ^kN)+2\sqrt{\epsilon N}K^i{}_{j}{}^{(3)}\nabla _iN^j
       \nonumber \\
       & & +2\sqrt{\epsilon N}N^{j}{}^{(3)}\nabla _k K^k{}_j-N^j \sqrt{
       \epsilon N}{}^{(3)}\nabla _j K^k{}_k+\epsilon N K_{kj}K^{kj}
       \nonumber \\
       & & +N^lN^j{}^{(3)}R_{lj}+\epsilon N^lN^jK_{lj}K^k{}_k-2\epsilon N^iN^k
       K_{kj}K^j{}_i+\epsilon \frac{N^iN^k}{\sqrt{\epsilon N}}
       \stackrel{\mbox{\LARGE .}}{K}_{ki}
       \nonumber \\
       & & -\frac{N^iN^k}{2N}{}^{(3)}\nabla _k(\partial _i N)-2\frac{\epsilon}
       {\sqrt{\epsilon N}}N^iN^kK_{kj}{}^{(3)}\nabla _iN^j
       \nonumber \\
       & &-\frac{\epsilon}{\sqrt{\epsilon N}}N^iN^jN^k{}^{(3)}\nabla _j K_{ik}
       -\frac{1}{4N}\partial _k N\partial^kN
       \nonumber \\
       & &+\frac{1}{4N^2}N^iN^k\partial _iN\partial _kN
\end{eqnarray}
We note that the last two terms in the expression of $R_{00}$ are not present
in the analogous formula without change of signature (see the appendix). The
other terms are, in general, slighty modified in comparison with the usual
formula (typically, $N$ is replaced by $\sqrt{\epsilon N}$). However, the
term $\epsilon$ appears sometimes outside the square root of the lapse
function and we shall see that the consequences of this simple fact will be
very important. The expression giving the ($0i$)-components of the Ricci
tensor is:
\begin{eqnarray}
\label{20}
R_{0i} &=& \frac{\epsilon}{\sqrt{\epsilon N}}N^k
       \stackrel{\mbox{\LARGE .}}{K}_{ki} +N^j{}^{(3)}
       R_{ij}-\frac{N^k}{2N}{}^{(3)}\nabla _k(\partial _i N)-\sqrt{\epsilon
       N}\partial _iK^j{}_j
       \nonumber \\
       & & +\sqrt{\epsilon N} {}^{(3)}\nabla _jK^j{}_i+\epsilon N^kK_{ik}
       K^j{}_j-\frac{\epsilon}{\sqrt{\epsilon N}}N^jN^m{}^{(3)}\nabla _j
       K_{im}
       \nonumber \\
       & &-\frac{\epsilon}{\sqrt{\epsilon N}}N^kK_{jk}{}^{(3)}\nabla _iN^j
       -\frac{\epsilon}{\sqrt{\epsilon N}}N^jK_{im}{}^{(3)}\nabla _jN^m
       -2\epsilon N^jK_i{}^mK_{mj}
       \nonumber \\
       & &+\frac{1}{4N^2}N^k\partial _kN\partial _iN
\end{eqnarray}
As in the previous case, we note that the last term is absent in the
usual expression. The space-space components of the Ricci tensor can be
written as:
\begin{eqnarray}
\label{21}
R_{ij} &=& {}^{(3)}R_{ij}+\frac{\epsilon}{\sqrt{\epsilon N}}
       \stackrel{\mbox{\LARGE .}}{K}_{ij}-
       2\epsilon K_{ki}K^k{}j+\epsilon K^k{}_kK_{ij}
       \nonumber \\
       & & -\epsilon \frac{N^k}{\sqrt{\epsilon N}}{}^{(3)}\nabla _kK_{ij}
       -\frac{1}{2N}{}^{(3)}\nabla _j(\partial _iN)-\frac{\epsilon}{
       \sqrt{\epsilon}N}K_{ki}{}^{(3)}\nabla _j N^k
       \nonumber \\
       & &-\frac{\epsilon}{\sqrt{\epsilon N}}K_{jk}{}^{(3)}\nabla _iN^k
       +\frac{\partial _iN\partial _jN}{4N^2}
\end{eqnarray}
Here again, an extra term (the last one) is present. Then, using the expression
of the inverse metric, we can compute the Ricci scalar; after quite
straightforward calculations, we find the following expression:
\begin{eqnarray}
\label{22}
R &=& {}^{(3)}R-3\epsilon K_{ki}K^{ki}+\epsilon K^2-2\frac{\epsilon}
  {\sqrt{\epsilon N}}N^i\partial _iK^j{}j
  \nonumber \\
  & & +2\frac{\epsilon}{\sqrt{\epsilon N}}h^{ki}
  \stackrel{\mbox{\LARGE .}}{K}_{ki}-\frac{1}{N}
  {}^{(3)}\nabla _k(\partial ^kN)-\frac{4\epsilon}{\sqrt{\epsilon N}}
  K_k{}^j{}^{(3)}\nabla _jN^k
  \nonumber \\
  & & +\frac{1}{2N^2}\partial _kN\partial ^k N
\end{eqnarray}
Some of the additional terms have been eliminated, but an extra term still
remains. Noting that:
\begin{equation}
\label{23}
h^{ki}\stackrel{\mbox{\LARGE .}}{K}_{ki}=\stackrel{\mbox{\LARGE .}}{K}
+2\sqrt{\epsilon N}K^{ki}K_{ki}+2K_{ki}{}^{(3)}\nabla ^kN^i
\end{equation}
we can re-express the formula for the Ricci scalar curvature; it takes the
following form:
\begin{eqnarray}
\label{24}
R &=& {}^{(3)}R+\epsilon K_{ki}K^{ki}+\epsilon K^2+\frac{2\epsilon}{\sqrt{
  \epsilon N}}\stackrel{\mbox{\LARGE .}}{K}-\frac{2\epsilon}
  {\sqrt{\epsilon N}}N^i\partial _i K
  \nonumber \\
  & &-\frac{1}{N}{}^{(3)}\nabla _k(\partial ^k N)+\frac{1}{2N^2}\partial _kN
  \partial ^kN
\end{eqnarray}
Now, we can compute the Einstein-Hilbert Lagrangian with change
of signature, which is
\begin{equation}
\label{25}
{\cal L}_G = \sqrt{-g}R=h^{\frac{1}{2}}\sqrt{\epsilon N}R
\end{equation}
Using the previous expression for R, we obtain the following definition of
the Lagrangian of the gravitational field:
\begin{eqnarray}
\label{27}
{\cal L}_G &=& h^{\frac{1}{2}}\sqrt{\epsilon N}({}^{(3)}R+\epsilon K_{ki}
           K^{ki}-\epsilon K^2) \nonumber \\
           & & +2\epsilon \frac{{\rm d}}{{\rm d}t}(h^{\frac{1}{2}}K)-2
           \partial _i(\epsilon h^{\frac{1}{2}}KN^i+h^{\frac{1}{2}}h^{ki}
           \partial _k \sqrt{\epsilon N})
\end{eqnarray}
This equation looks like the usual one since it contains three parts, one
being composed by the kinetic and  potential terms, the two others
representing surface terms. However, we note the presence of the factor
$\epsilon $ in the dynamical term. The extra terms have disappeared or have
been incorporated in the surface terms which are slighty modified in
comparison with the case without the change of signature. In what follows, we
shall drop out the surface terms and shall consider only the Lagrangian
${\cal L}_G$ defined by:
\begin{equation}
\label{28}
{\cal L}_G[N, N^i, h_{ij}]=h^{\frac{1}{2}}\sqrt{\epsilon N}({}^{(3)}R+
\epsilon K_{ij}K^{ij}-\epsilon K^2)
\end{equation}
Starting from the previous equation, we can perform now full Hamiltonian
analysis. The conjugate momenta are given by:
\begin{eqnarray}
\label{29}
\pi_{\mu} &=& \frac{{\rm \delta } {\cal L}_G}{{\rm \delta } (\partial _tN^
{\mu})}=0 \\
\label{30}
\pi^{ij} &=& \epsilon h^{\frac{1}{2}}(K^{ij}-h^{ij}K)
\end{eqnarray}
where $K=h^{ij}K_{ij}$. Equation (\ref{29}) shows that ${\cal L}_G$ is a
constrained Lagrangian and therefore requires the use of the Dirac formalism
developed especially to treat this kind of systems \cite{r6,r7,r8,r9}.
Equation (\ref{30}) can be inversed to provide the following expression for
$K^{ij}$:
\begin{equation}
\label{31}
K^{ij}=\epsilon h^{-\frac{1}{2}}(\pi^{ij}-\frac{\pi}{2}h^{ij})
\end{equation}
where $\pi=h^{ij}\pi_{ij}$. The canonical hamiltonian can be written as:
\begin{eqnarray}
\label{32}
{\cal H}_G &=& \pi_{\mu}\partial _tN^{\mu}+\pi^{ij}\partial _th_{ij}-{\cal L}
           _G \\
\label{33}
           &=&2\pi^{ij}{}^{(3)}\nabla _iN_j-h^{-\frac{1}{2}}\sqrt{\epsilon N}
           (\frac{\epsilon}{2}\pi ^2-\epsilon \pi^{ij}\pi_{ij}+h{}^{(3)}R)
\end{eqnarray}
Integrating by parts and dropping out the surface term, we obtain:
\begin{equation}
\label{34}
H_c=\int {\rm d}^3x(\sqrt{\epsilon N}{\cal H}+N_j{\cal H}^j)
\end{equation}
where the expressions of ${\cal H}$ and ${\cal H}^j$ are:
\begin{eqnarray}
\label{35}
{\cal H} &=& h^{-\frac{1}{2}}(\epsilon \pi ^{ij}\pi_{ij}-\frac{\epsilon}{2}
         \pi ^2)-h^{\frac{1}{2}}{}^{(3)}R \\
\label{36}
{\cal H}^j &=& -2{}^{(3)}\nabla _i\pi ^{ij}
\end{eqnarray}
${\cal H}$ can be written in such a way that the metric of the superspace
appears explicitly:
\begin{equation}
\label{37}
{\cal H}=G_{ijkl}\pi ^{ij}\pi^{kl}-h^{\frac{1}{2}}{}^{(3)}R
\end{equation}
where $G_{ijkl}$ is defined by:
\begin{equation}
\label{38}
G_{ijkl}=\frac{\epsilon}{2}h^{-\frac{1}{2}}(h_{ik}h_{jl}+h_{il}h_{jk}-
h_{ij}h_{kl})
\end{equation}
We see that the only effect of the change of signature is to modify the
respective sign between the kinetic part and the potential part in the
definition of the Hamiltonian (\ref{37}). In other words, the change
from an Euclidean signature to a Lorentzian signature in the physical
manifold $M$ induces a change in the superspace from the signature ($-+++++$)
to the signature ($+-----$). We note that the superspace metric remains
globally hyperbolic even if the metric of $M$ happens to be Euclidean. This
becomes more explicit if we define the coordinate $\xi$ in the superspace
following \cite{r10}:
\begin{equation}
\label{39}
\xi=4\sqrt{\frac{2}{3}}h^{\frac{1}{4}}
\end{equation}
and five coordinates $\xi ^A$ ($A=1\ldots 5$) orthogonal (in the sens of the
supermetric) to $\xi$. Then, the interval in the superspace takes on
the form:
\begin{equation}
\label{40}
{\rm d}s^2=-\epsilon{\rm d}\xi ^2+\frac{3}{8}\epsilon \stackrel{-}{G}_{AB}
{\rm d}\xi ^A{\rm d}\xi ^B
\end{equation}
where $\stackrel{-}{G}_{AB}$ is the metric of a five-dimensional subspace
of the superspace whose signature
is ($+++++$). Using previous relations, we can compute the curvature
invariants ($R_{ABCD}R^{ABCD}$ for example) and the geodesics in the
superspace \cite{r10} and show that all these quantities are well-behaving
when the change of signature from ($-+++++$) to ($+-----$) occurs. This
suggests another argument in favor of considering the solutions of Einstein's
equations with change of signature, since there exist a priori no reason for
studying only a limited region of the superspace.
\par
In the next section, we are going to consider minisuperspaces and to show
how the equations of motion can be found using the Dirac formalism for
constrained systems.

\section{Applications to minisuperspaces}

\setcounter{equation}{0}
In this section, we restrict our considerations to minisuperspaces \cite{r11}.
This means that we impose very strong symmetry conditions on the solutions
so that almost all degrees of freedom are frozen, except a few ones, which
leads to the superspace of finite dimension in which the analytic
computations and finite integrations can be performed. The coordinates
$q^{\alpha}$ in the minisuperspace are the components of the metric tensor
and eventually the fields describing the matter. In the case of change of
the signature, the Lagrangian can be written as:
\begin{equation}
\label{41}
L=\sqrt{\epsilon N}\biggl(\frac{1}{2\epsilon N}f_{\alpha \beta}
\stackrel{\mbox{\LARGE .}}{q}^{\alpha}\stackrel{\mbox{\LARGE .}}{q}^{\beta}
-V(q^{\alpha})\biggr)
\end{equation}
where $f_{\alpha \beta}$ is the metric $G_{ijkl}$ restricted to the
minisuperspace, and the canonical Hamiltonian deduced from the above
expression of the Lagrangian is given by:
\begin{equation}
\label{42}
H_c=\sqrt{\epsilon N}\biggl(\frac{1}{2}f^{\alpha \beta}\pi_{\alpha}\pi_{\beta}
+V(q^{\alpha})\biggr)
\end{equation}
In what follows, we shall suppose that the manifold has the geometry of
a Friedmann-Robertson-Walker (FRW) Universe:
\begin{equation}
\label{43}
{\rm d}s^2=-N(t){\rm d}t^2+a^2(t)\biggl(\frac{{\rm d}r^2}{1-kr^2}+
r^2({\rm d}\theta ^2+\sin ^2\theta {\rm d}\varphi ^2)\biggr)
\end{equation}
where $k$ can take on the values $0,\pm 1$ according to the case where the
three-dimensional spacelike hypersurfaces are flat, open or closed. The
matter will be described by a unique scalar field. Putting equation
(\ref{43}) in the expression (\ref{28}) of the Lagangian, we find that the
action for the gravitational field can be written as:
\begin{equation}
\label{44}
S_G=\alpha \int {\rm d}t (k\sqrt{\epsilon N}a-\epsilon
\frac{\stackrel{\mbox{\LARGE .}}{a}^2a}{\sqrt{\epsilon N}})
\end{equation}
where $\alpha$ is a dimensional constant. With the assumption that the scalar
field is spatially homogeneous, that is to say $\phi =\phi (t)$, the action
of the matter takes on the form:
\begin{eqnarray}
\label{45}
S_M &=& -\int {\rm d}^4x\sqrt{\epsilon N}h^{\frac{1}{2}}\biggl(\frac{1}{2}
    g^{\mu \nu}\partial _{\mu}\phi \partial _{\nu}\phi+V(\phi)\biggr) \\
    &=&\frac{\alpha}{6}\int {\rm d}t a^3\biggl(
    \frac{\epsilon \stackrel{\mbox{\LARGE .}}{\phi }^2}{2
    \sqrt{\epsilon N}}-\sqrt{\epsilon N}V(\phi)\biggr)
\end{eqnarray}
The full Lagrangian of the system (gravitational field and scalar field),
$L=L_G+L_M$ is given by:
\begin{equation}
\label{46}
L=\sqrt{\epsilon N}\biggl(\frac{1}{2\epsilon N}(-2\epsilon a
\stackrel{\mbox{\LARGE .}}{a}^2+\epsilon \frac{a^3}{6}
\stackrel{\mbox{\LARGE .}}{\phi }^2)-U(a,\phi)\biggr)
\end{equation}
where the potential $U(a,\phi)$ is defined by:
\begin{equation}
\label{47}
U(a,\phi)=\frac{a^3}{6}V(\phi)-ka
\end{equation}
The metric $f_{\alpha \beta}$ of the minisuperspace can readily be obtained
from the expression of $L$:
\[ f_{\alpha \beta}=\left( \begin{array}{clcr}
   -\frac{\epsilon}{2a} & 0 \\
   0 & \frac{6\epsilon}{a^3}
   \end{array}
\right) \]
As announced, the signature of the metric of the minisuperspace changes from
($-+$) to ($+-$) when the signature of the FRW metric becomes Lorentzian.
We can now compute the conjugate momenta:
\begin{eqnarray}
\label{48}
\pi _a &=& \frac{{\rm \partial}L}{{\rm \partial}
       \stackrel{\mbox{\LARGE .}}{a}}=-2\epsilon \frac{
       \stackrel{\mbox{\LARGE .}}{a}a}{\sqrt{\epsilon N}} \\
\label{49}
\pi _{\phi} &=& \frac{{\rm \partial}L}{{\rm \partial}
            \stackrel{\mbox{\LARGE .}}{\phi }}=\frac{\epsilon a^3}
            {6\sqrt{\epsilon N}}\stackrel{\mbox{\LARGE .}}{\phi } \\
\label{50}
\pi _N &=& 0
\end{eqnarray}
The last equality is a primary constraint and is a consequence of the gauge
invariance of General Relativity (Thus, following Dirac's classification, it
is also a {\em first class} constraint). The canonical Hamiltonian is given
by:
\begin{eqnarray}
\label{51}
H_c &=& \sqrt{\epsilon N}\biggl(-\frac{\epsilon }{4a}\pi _a^2+\frac{3\epsilon
}{
    a^3}\pi _{\phi}^2+U(a,\phi)\biggr)\\
\label{52}
    &=& \sqrt{\epsilon N}H
\end{eqnarray}
One can easily verify that $H_c$ has the form given by the equation
(\ref{42}). Then, the total Hamiltonian can be written as:
\begin{equation}
\label{53}
H_T=\sqrt{\epsilon N}H+\lambda \pi _N
\end{equation}
where $\lambda$ is a Lagrange multiplier. According to the Dirac algorithm
for constrained systems, we have to check that the primary constraint
$\pi _N\approx 0$ ($\approx$ is the Dirac weak equality) is preserved in time:
\begin{equation}
\label{54}
\{\pi _N, H_T\}_P\approx 0
\end{equation}
where $\{ \ \ , \ \  \}_P$ is the Poisson bracket in the minisuperspace
defined by \footnote{This the restricted version to the minisuperspace of the
general
formula:
\[
\{f, g\}_P=\sum _{i,j}\int {\rm d}^3x(\frac{{\rm \delta}f(z)}{{\rm \delta}
h_{ij}(x)}\frac{{\rm \delta}g(z)}{{\rm \delta}\pi ^{ij}(x)}
-\frac{{\rm \delta}f(z)}{{\rm \delta}\pi ^{ij}(x)}\frac{{\rm \delta}g(z)}
{{\rm \delta}h_{ij}(x)})
\nonumber
\]
}:
\begin{equation}
\label{56}
\{f, g\}_P=\frac{{\rm \partial }f}{{\rm \partial }N}\frac{{\rm \partial }g}
{{\rm \partial }\pi _N}+\frac{{\rm \partial }f}{{\rm \partial }a}\frac{{\rm
\partial }g}{{\rm \partial }\pi _a}+\frac{{\rm \partial }f}
{{\rm \partial }\phi }
\frac{{\rm \partial }g}{{\rm \partial }\pi _{\phi }}-
\frac{{\rm \partial }f}{{\rm \partial }\pi _N}\frac{{\rm \partial }g}
{{\rm \partial }N}-\frac{{\rm \partial }f}{{\rm \partial }\pi _a}
\frac{{\rm \partial }g}{{\rm \partial }a}-\frac{{\rm \partial }f}
{{\rm \partial }\pi _{\phi }}\frac{{\rm \partial }g}{{\rm \partial }\phi }
\end{equation}
This leads to the secondary constraint:
\begin{equation}
\label{57}
-\frac{\epsilon}{2\sqrt{\epsilon N}}H\approx 0
\end{equation}
We note the difference in comparison with the usual case where we obtain
$H\approx 0$. One have to be very carefull since equation (\ref{57}) has the
form $f(p, q)g(p, q)\approx 0$. Earliest works \cite{r12} have shown that if
we assume that the last relation implies $f(p, q)\approx 0$ or $g(p, q)
\approx 0$, this can lead to a contradiction with the Dirac conjecture which
states that all secondary first class constraints must generate symmetries.
Therefore, at this stage, we are not allowed to deduce $H\approx 0$ from
equation (\ref{57}). On the other hand, if the relation
$h(p, q)\approx 0$ holds, it implies that for all functions $f(p, q)$,
the equation $f(p, q)h(p, q)\approx 0$ must hold, too. For consistency,
the secondary constraints must be preserved in time:
\begin{eqnarray}
\label{58}
\{-\frac{\epsilon}{2\sqrt{\epsilon N}}H, H_T\}_P &=& -\frac{\epsilon \lambda}
{2}\{\frac{1}{\sqrt{\epsilon N}}, \pi _N\}_PH \\
\label{59}
&=& \frac{\lambda }{2\epsilon N^2}(-\frac{\epsilon}{2\sqrt{\epsilon N}}H) \\
\label{60}
&\approx& 0
\end{eqnarray}
according to the previous discussion. Thus, there are no further constraints.
The two constraints are first class ones since their Poisson bracket vanishes.
The extended Hamiltonian can be written now as:
\begin{equation}
\label{61}
H_E=\lambda \pi _N+H(\sqrt{\epsilon N}-\frac{\eta \epsilon}
{2\sqrt{\epsilon N}})
\end{equation}
where $\eta$ is a Lagrange multiplier. We can note that this expression is
different from the usual one (i.e $H_E=\lambda \pi _N+NH$). Finally, we want
to show how the equations of motion can be deduced from the extended
Hamiltonian.
The Hamilton equations are:
\begin{equation}
\label{62}
\stackrel{\mbox{\LARGE .}}{N}\approx \{N, H_E\}_P=\lambda
\end{equation}
This equation shows that $N$ is completely arbitrary since its derivative is
equal to the Lagrange multiplier. Next,
\begin{eqnarray}
\label{63}
\stackrel{\mbox{\LARGE .}}{\pi _N} &\approx &\{\pi _N, H_E\}_P  \\
\label{64}
             &\approx &-\frac{\epsilon }{2\sqrt{\epsilon N}}H+\frac{H\eta}
             {2N}(-\frac{\epsilon}{2\sqrt{\epsilon N}}H) \\
\label{65}
             &=& 0
\end{eqnarray}
where we have used the constraints to transform the weak equality into a
strong one.
\begin{eqnarray}
\label{66}
\stackrel{\mbox{\LARGE .}}{a} &\approx & \{a, H_E\}_P \\
\label{67}
\stackrel{\mbox{\LARGE .}}{a} &\approx & (\sqrt{\epsilon N}-\frac{\eta
\epsilon }{2\sqrt{\epsilon N}})\{a, H\}_P \\
\label{68}
        &\approx &-\frac{1}{\sqrt{\epsilon N}}(N-\frac{\eta}{2})\frac{
        \pi _a}{2a}
\end{eqnarray}
But as we have just seen with expression (\ref{62}), $N$ is arbitrary.
Therefore, the last equation can be written as:
\begin{eqnarray}
\label{69}
\stackrel{\mbox{\LARGE .}}{a} &\approx &-\frac{N}{\sqrt{\epsilon N}}\frac{\pi
_a}{2a} \\
\label{70}
        &\approx &-\epsilon \sqrt{\epsilon N}\frac{\pi _a}{2a}
\end{eqnarray}
The equation of motion for $\pi _a$ is given by:
\begin{eqnarray}
\label{71}
\stackrel{\mbox{\LARGE .}}{\pi _a} &\approx & \{\pi _a, H_E\}_P \\
\label{72}
             &\approx & (\sqrt{\epsilon N}-\frac{\eta \epsilon}{2\sqrt{
             \epsilon N}})\{\pi _a, H\}_P \\
\label{73}
             &\approx & \sqrt{\epsilon N}(-\frac{\epsilon}{4a^2}\pi _a^2+
             \frac{9\epsilon }{a^4}\pi _{\phi}^2+k-\frac{a^2}{2}V(\phi ))
\end{eqnarray}
For the scalar field, we obtain:
\begin{eqnarray}
\label{74}
\stackrel{\mbox{\LARGE .}}{\phi } &\approx & \{\phi, H_E\}_P \\
\label{75}
            &\approx & (\sqrt{\epsilon N}-\frac{\eta \epsilon}{2
            \sqrt{\epsilon N}})\{\phi, \frac{3\epsilon}{a^3}\pi _{\phi}^2\}_P
\\
\label{76}
            &\approx & \epsilon \sqrt{\epsilon N}\frac{6}{a^3} \pi _{\phi}
\end{eqnarray}
The dynamical equation for the conjugate momentum of the scalar field is:
\begin{eqnarray}
\label{77}
\stackrel{\mbox{\LARGE .}}{\pi }_{\phi} &\approx & \{\pi _{\phi}, H_E\}_P \\
\label{78}
                  &\approx & -\sqrt{\epsilon N}\frac{a^3}{6}\frac{
                  {\rm d}V}{{\rm d}\phi}
\end{eqnarray}
It is now straightforward to show that the previous equations are equivalent
to the equations already found in earlier papers \cite{r1,r2} describing
the change of signature of a FRW Universe filled with a scalar field:
\begin{eqnarray}
\label{79}
\frac{\stackrel{\mbox{\LARGE ..}}{a}}{a}
-\frac{\stackrel{\mbox{\LARGE .}}{a}\stackrel{\mbox{\LARGE .}}{N}}{2aN}
+\frac{\stackrel{\mbox{\LARGE .}}{\phi }^2}{8} &-&\frac{N}{6}V = 0 \\
\label{80}
\stackrel{\mbox{\LARGE ..}}{\phi }
-\frac{\stackrel{\mbox{\LARGE .}}{N}\stackrel{\mbox{\LARGE .}}{\phi }}{2N}
+3\frac{\stackrel{\mbox{\LARGE .}}{a}}{a}\stackrel{\mbox{\LARGE .}}{\phi }
&+& N\frac{{\rm d}V}{{\rm d}\phi} = 0 \\
\label{81}
\frac{kN}{a^2}+\frac{\stackrel{\mbox{\LARGE .}}{a}^2}{a^2}=
\frac{N}{6}\biggl(\frac{\stackrel{\mbox{\LARGE .}}{\phi }^2}{2N}
&+& V(\phi )\biggr)
\end{eqnarray}
Thus, we have verified that the hamiltonian formalism adapted to the case
including the change of signature, used with care, provides the right
equations of motion. Obviously, the choice of $N$ remains to be made in
order to obtain a good junction condition. This point has been discussed
with more details in Refs. \cite{r1,r2,r12.1,r12.2}.

\section{Canonical quantization}

\setcounter{equation}{0}
Following the suggestion of Hartle and Hawking \cite{r13}, we shall suppose
that an Euclidean region could have existed in the very early Universe.
In this regime, the behaviour of the wave function of the Universe
corresponds to the behaviour of a wave function of usual quantum mechanics in
a classically forbidden region. However, we have seen that classical
Euclidean solutions of Einstein's equations can also exist, so that the
problem of quantization of such solutions and of the behaviour of the
wave function in the Euclidean regime naturally arises. Let us study this
problem now.
\par
We want to quantize a theory with the constraints of first class described
by the extended Hamiltonian \cite{r14}:
\begin{equation}
\label{82}
H_E=\lambda \pi _N+(\sqrt{\epsilon N}-\frac{\eta \epsilon}{2
\sqrt{\epsilon N}})H
\end{equation}
According to the prescriptions of Dirac, this leads to the relations:
\begin{equation}
\label{83}
\hat{\pi }_N\Psi (a, \phi , N)=0\ \ \Rightarrow \ \ \Psi=\Psi (a, \phi )
\end{equation}
\begin{equation}
\label{84}
(\sqrt{\epsilon \hat{N}}-\frac{\eta \epsilon }{2\sqrt{\epsilon \hat{N}}})
\hat{H}\Psi(a, \phi)=0\ \ \Rightarrow \ \ \hat{H}\Psi (a, \phi )=0,
\end{equation}
the last equation resulting from the fact that the wave function
$\Psi (a, \phi )$ does not depend on $N$ (see the equation (\ref{83})).
Consequently, the Wheeler-De Witt equation is given by \cite{r11}:
\begin{equation}
\label{85}
(-\frac{\epsilon}{4a}\pi _a^2+\frac{3\epsilon }{a^3}\pi _{\phi}^2-ka+
\frac{a^3V(\phi)}{6})\Psi (a, \phi)=0
\end{equation}
where the hat on the symbols denoting the operators has been omitted.
If we multiply by $-\epsilon a$, we obtain:
\begin{equation}
\label{86}
\biggl(\frac{1}{4}\pi _a^2-\frac{3}{a^2}\pi _{\phi}^2-\epsilon a^2(a^2\frac{
V(\phi )}{6}-k)\biggr)\Psi (a, \phi )=0
\end{equation}
Now, we can interpret the change of signature in a different way: the last
equation has the usual Wheeler-De Witt form (no change from
($-+$) to ($+-$)), but the potential is modified by the factor $\epsilon $
which takes into account the difference of signs between the Euclidean and
Lorentzian regions:
\begin{equation}
\label{87}
\tilde{U}(a, \phi)=\epsilon a^2(a^2\frac{V(\phi )}{6}-k)
\end{equation}
In other words, all the effects of the signature change have been
incorporated into the potential term only. Because of the presence of factor
$\epsilon $, we have to check if $\tilde{U}(a, \phi)$ is
well-behaving at the surface of change $\Sigma$. If we assume that
$N(t)=\epsilon $, then the equation (\ref{81}) becomes:
\begin{equation}
\label{88}
(\frac{\stackrel{\mbox{\LARGE .}}{a}}{a})^2=\frac{
\stackrel{\mbox{\LARGE .}}{\phi }^2}{12}+\epsilon (\frac{V}{6}-\frac{k}{a^2})
\end{equation}
If we require the continuity of the scale factor and of its first derivative,
this provides a criterion (cf. \cite{r1}, \cite{r2}) for determining the
value of $V$ at which the change of signature does occur:
\begin{eqnarray}
\label{89}
V &=& \frac{6k}{a^2} \ \ \ \mbox{on $\Sigma$} \\
\label{90}
V &>& \frac{6k}{a^2} \ \ \ \mbox{if $\epsilon =1$} \\
\label{91}
V &<& \frac{6k}{a^2} \ \ \ \mbox{if $\epsilon =-1$}
\end{eqnarray}
This criterion assures that the potential $\tilde{U}(a, \phi)$ is continuous
(but not its derivative) when the change of signature occurs. Equation
(\ref{86}) leads to a differential equation if we apply the correspondence
principle, that is to say when we replace the square of the conjugate
momentum according to the usual rule \cite{r15}:
\begin{equation}
\label{92}
\pi ^2\ \ \longrightarrow \ \ \ -q^{-p}\frac{{\rm \partial }}{{\rm \partial }q}
(q^p\frac{{\rm \partial }}{{\rm \partial }q})
\end{equation}
where the exponent $p$ takes into account the operator-ordering problem. In
what follows, we shall assume that $p=-1$ in order to obtain a differential
equation which can be solved analytically. We shall also assume that the
scalar field (and therefore the potential $V(\phi )$) is constant and plays
the r\^ ole of a cosmological term in Einstein's equations. This will enable us
to find a connection with the classical solutions displaying a change of
signature already found in \cite{r1} and \cite{r2}.
With these assumptions, the Wheeler-De Witt equation can be written as:
\begin{equation}
\label{93}
\frac{{\rm d}^2}{{\rm d}a^2}\Psi (a)-\frac{1}{a}\frac{{\rm d}}{{\rm d}a}
\Psi (a)+4\epsilon a^2(H^2a^2-1)\Psi (a)=0
\end{equation}
where $H^2=\frac{V}{6}$ and $k=1$. By defining $z=-\epsilon H^{-\frac{4}{3}}
(H^2a^2-1)$, the above equation becomes:
\begin{equation}
\label{94}
\frac{{\rm d}^2}{{\rm d}z^2}\Psi (z)-z\Psi (z)=0
\end{equation}
which is the differential equation defining the Airy functions \cite{r16,r17}.
The general solution of the Wheeler-De Witt equation is given in this case
by:
\begin{equation}
\label{95}
\Psi (z)=\lambda (\phi )Ai(z)+\eta (\phi )Bi(z)
\end{equation}
where $Ai(z)$ and $Bi(z)$ are the Airy functions of first kind and second
kind. $\lambda $ and $\eta $ are arbitrary functions of the parameter
$\phi $. When the scale factor $a(t)$ comes close to zero, the term $a^4$
can be neglected and the Wheeler-De Witt equation can be written as:
\begin{equation}
\label{96}
\frac{{\rm d}^2\Psi}{{\rm d}a^2}-\frac{1}{a}\frac{{\rm d}\Psi }{{\rm d}a}
+4a^2\Psi=0
\end{equation}
The condition $a(t)\longrightarrow 0$ implies that $\epsilon $ must
be equal to $-1$. Then, $\Psi(a=0, \phi)$ is a constant and does not depend
on $\phi $. Since $z(a=0)=-H^{-\frac{4}{3}}$, the general solution (\ref{95})
is (in what follows we shall put a ``CS'' superscript meaning ``Change of
Signature" to distinguish the wave functions obtained here from the wave
functions of usual quantum cosmology):
\begin{equation}
\label{97}
\Psi ^{(CS)}(z)=\frac{\alpha Ai\biggl(-\epsilon
H^{-\frac{4}{3}}(H^2a^2-1)\biggr ) +
\beta Bi\biggl(-\epsilon H^{-\frac{4}{3}}(H^2a^2-1)\biggr )}{\alpha
Ai(-H^{-\frac{4}{3}})+\beta Bi(-H^{-\frac{4}{3}})}
\end{equation}
In order to determine the value of $\alpha $ and $\beta $, we have to
solve the problem of boundary conditions of quantum cosmology.
\par
Let us analyze the difference between the wave functions (\ref{97}) and those
of usual quantum cosmology in each region separately (i.e. for
$\epsilon =\pm 1$). In the region where $a<\frac{1}{H} (\epsilon =-1)$, the
WKB solutions of the Wheeler-De Witt equation take on the form:
\begin{equation}
\label{98}
\Psi ^{(CS)}\propto e^{\pm i\biggl(\frac{2}{3H^2}(1-H^2a^2)^{\frac{3}{2}}
+\frac{\pi }{4}\biggr)}
\end{equation}
In the simplified model considered here, the wave function of the Universe
is formally equivalent to a wave function of a particle with energy equal
to zero \cite{r11}. Since for $a<\frac{1}{H}$, the potential of the usual
Wheeler-De Witt equation is positive, this region is classically forbidden
and we obtain an exponential behaviour $\Psi \sim e^{-I}$, where $I$ denotes
the Euclidean action: this phenomenon is the well-known tunnel effect. The
fact that the wave function $\Psi ^{(CS)}$ adopts an oscillatory behaviour
seems quite natural here since, due to the value $-1$ of the term
$\epsilon $, the potential is now negative and the energy of the fictitious
particle is above this value, so that the Euclidean region is no longer
forbidden. Using Wigner's functions \cite{r18,r19,r20,r21,r22} or canonical
transformations \cite{r23} it is easy to show that oscillatory behaviour
means that $\Psi ^{(CS)}$ carries quantum correlations. Obviously, these
correlations correspond to the classical solutions of Einstein's equations
with change of signature.
\par
In the region $a>\frac{1}{H} (\epsilon =1)$, the wave function is supposed
to describe our Universe and we can use it for cosmological interpretations.
The predictions coming from $\Psi ^{(CS)}$ should be compared with those
coming from the wave functions of ordinary quantum cosmology (written without
``CS'' superscript). To fix the wave function, we have to
solve the problem of boundary conditions. Many propositions have
been made, one of the best known is the ``no boundary'' proposition of
Hawking and Hartle \cite{r13}. The wave function on the three-dimensional
hypersurface $B$ is constructed by performing the Euclidean path integral:
\begin{equation}
\label{99}
\Psi [\tilde{h_{ij}}, \tilde{\Phi}, B]=\sum _{M}\int {\cal D}g_{\mu \nu}
{\cal D}\Phi \exp (-I[g_{\mu \nu}, \Phi])
\end{equation}
where the sum is taken over all manifolds $M$ having $B$ as part of their
boundary and over all metrics $g_{\mu \nu}$ and matter fields $\Phi $ which
induce $\tilde{h_{ij}}$ and $\tilde{\Phi }$ on $B$; $I$ denotes the Euclidean
action. The proposal of Hawking and Hartle consists in restricting the sum
to compact four-dimensional manifolds $M$ whose only boundary is $B$.
In the case of the minisuperspace considered, this would lead to a
wave function which takes on the form ($\alpha =1$, $\beta =0$):
\begin{equation}
\label{100}
\Psi _{HH}(z)=\frac{Ai\biggl(H^{-\frac{4}{3}}(1-H^2a^2)\biggr)}
{Ai(H^{-\frac{4}{3}})}
\end{equation}
We don't know how to compute $\Psi ^{(CS)}_{HH}$. Another well-known choice
is the tunneling boundary condition of Vilenkin \cite{r24}. The wave function
is assumed to have only a WKB component $e^{-iS}$ in the semi-classical
regime. This choice is made in order to describe the birth of the Universe
that might be interpreted as the tunnel effect of common quantum mechanics.
This leads to the following wave functions ($\alpha =1$, $\beta =i$):
\begin{eqnarray}
\label{101}
\Psi _{V}(z) &=& \frac{Ai\biggl( H^{-\frac{4}{3}}(1-H^2a^2)\biggr)+
iBi\biggl( H^{-\frac{4}{3}}(1-H^2a^2)\biggr)}{Ai(H^{-\frac{4}{3}})+iBi(H^
{-\frac{4}{3}})} \\
\label{102}
\Psi ^{(CS)}_{V}(z) &=& \frac{Ai\biggl( H^{-\frac{4}{3}}(1-H^2a^2)\biggr)+
iBi\biggl( H^{-\frac{4}{3}}(1-H^2a^2)\biggr)}{Ai(-H^{-\frac{4}{3}})+iBi(-H^
{-\frac{4}{3}})}
\end{eqnarray}
$\Psi _{V}(z)$ and $\Psi ^{(CS)}_V(z)$ do not coincide, in the region where
$\epsilon =1$, only because of the presence of $-H^{-\frac{4}{3}}$ instead of
$H^{\frac{4}{3}}$ in the argument of the Airy functions of the denominator.
Clearly, this is true for all boundary conditions. It comes from the fact
that while computing the denominator, we used the condition $\Psi ^{(CS)}(
a=0, \phi )=0$ fixed in the region $\epsilon =-1$ where the classical
theories are not identical. As we shall see, this will have important
consequences. Another boundary condition seems also quite natural. It
consists in assuming the continuity of the first derivative of the wave
function at points at which the change of signature occurs:
\begin{equation}
\label{103}
\lim _{z\rightarrow 0^-}\Psi '_{\epsilon =-1}(z)=\lim _{z\rightarrow 0^+}
\Psi '_{\epsilon =+1}(z)
\end{equation}
We remind that it happens at $a=\frac{1}{H}$ or $z=0$. This condition leads
to:
\begin{equation}
\label{104}
-\alpha Ai'(0)-\beta Bi'(0)=\alpha Ai'(0)+\beta Bi'(0)
\end{equation}
Recalling one of the properties of the Airy functions which is \cite{r14,r15}:
\begin{equation}
\label{105}
Bi'(0)=-\sqrt{3}Ai'(0),
\end{equation}
we find that the wave function is now given by($\alpha =\sqrt{3}, \beta =1$):
\begin{equation}
\label{106}
\Psi ^{(CS)}_{C}(z)=\frac{\sqrt{3}Ai\biggl(H^{-\frac{4}{3}}(1-H^2a^2)\biggr)
+Bi\biggl(H^{-\frac{4}{3}}(1-H^2a^2)\biggr)}{\sqrt{3}Ai(-H^{-\frac{4}{3}})
+Bi(-H^{-\frac{4}{3}})}
\end{equation}
$\Psi ^{(CS)}_C$ is the only wave function which is of class $C^1$ when the
change of signature occurs. This boundary condition makes no sense in
ordinary quantum cosmology where $\Psi _C$ cannot appear (more precisely all
the wave functions of ordinary quantum cosmology are $C^1$-continuous). We
can write the WKB approximation $\Psi =Ae^{iS}$ for these four wave functions
using the asymptotic form of Airy's functions \cite{r16,r17}:
\begin{equation}
\label{107}
\lim_{z \rightarrow \infty} Ai(-z)=\frac{z^{-\frac{1}{4}}}{\sqrt{\pi }}\sin
(\frac{2}{3}z^{\frac{3}{2}}+\frac{\pi }{4})
\end{equation}
\begin{equation}
\label{108}
\lim_{z \rightarrow \infty} Bi(-z)=\frac{z^{-\frac{1}{4}}}{\sqrt{\pi }}\cos
(\frac{2}{3}z^{\frac{3}{2}}+\frac{\pi }{4})
\end{equation}
This asymptotic form can be used in the numerator of the expression
(\ref{97}) since $a\longrightarrow \infty $ in the region $\epsilon =1$ but
also in its denominator because $H^{-\frac{4}{3}}=(\frac{6}{V})^{\frac{2}{3}}
>>1$ since $|V|<<1$ in the semiclassical regime. We obtain (for
$\Psi^{(CS)}_C$ we have just considered the part of the wave function
corresponding to the Universe in expansion. One can show that the two
components $e^{iS}$ and $e^{-iS}$ does not interfere \cite{r27}):
\begin{eqnarray}
\label{109}
\Psi _{HH}(a, \phi) &\propto & e^{\frac{4}{V}}(H^2a^2-1)^{-\frac{1}{4}}
e^{-i\frac{2}{3H^2}(H^2a^2-1)^{\frac{3}{2}}} \\
\label{110}
\Psi ^{(CS)}_C(a, \phi ) &\propto & \frac{V^{-\frac{1}{3}}}
{\cos (\frac{4}{V}-\frac{\pi }{12})}(H^2a^2-1)^{-\frac{1}{4}}
e^{-i\frac{2}{3H^2}(H^2a^2-1)^{\frac{3}{2}}} \\
\label{111}
\Psi ^{(CS)}_V(a, \phi) &\propto & e^{i(\frac{4}{V}+\frac{\pi }{4})}
(H^2a^2-1)^{-\frac{1}{4}}e^{-i\frac{2}{3H^2}(H^2a^2-1)^{\frac{3}{2}}} \\
\label{112}
\Psi _V(a, \phi) &\propto & e^{-\frac{4}{V}}(H^2a^2-1)^{-\frac{1}{4}}
e^{-i\frac{2}{3H^2}(H^2a^2-1)^{\frac{3}{2}}}
\end{eqnarray}
Following Grishchuk and Gibbons \cite{r26,r27}, we can display $\Psi _{HH}$,
$\Psi ^{(CS)}_C$, $\Psi ^{(CS)}_V$ and $\Psi _V$ as seen in the space of all
the wave functions defined on our minisuperspace. Introducing the notation
$\alpha =|A|e^{i\gamma _1}$, $\beta =|B|e^{i\gamma _2}$, the choice of
boundary conditions is equivalent to fixing the complex number $\xi $
defined by:
\begin{equation}
\label{113}
\xi =xe^{i\gamma }
\end{equation}
where $x=\frac{|B|}{|A|}$ and $\gamma =\gamma _2-\gamma _1$. If we identify
$x$ with $\theta $ via the relation $x=cotan\frac{\theta }{2}$, the
numerator of the wave function is represented by a point $(\theta, \gamma)$ on
a sphere of unit radius. The denominator can be found by identifying its
modulus with the distance $\rho $ from the center of the sphere. Then a wave
function is characterized by a point $(\rho ,\theta ,\gamma )$ of a three-
dimensional parameter space (cf. the figure). For example, $\Psi ^{(CS)}_C$
corresponds to $\alpha =\sqrt{3}$ and $\beta =1$, then $\gamma =0$ and
$\theta =2\tan \sqrt{3}=\frac{2\pi }{3}$. Consequently $\Psi ^{(CS)}_C$ will
be represented on the radius ($\theta =\frac{2\pi }{3}, \gamma =0$).
$\Psi ^{(CS)}_V$ and $\Psi _V$ will be plotted in the same direction ($\theta
=\frac{\pi }{2}, \gamma =\frac{\pi }{2}$) but as their moduli differ, they
will be represented at different points. These wave functions are peaked
around the classical solutions:
\begin{eqnarray}
\label{114}
a(t) &=& \frac{1}{H}\cosh (Ht+t_0) \\
\label{115}
\phi &=& \phi _i
\end{eqnarray}
where the constant $\phi _i$ is the initial value of the scalar field and
$t_0$ a constant. Among these solutions there is the continuous solution,
corresponding to $t_0=0$. All the solutions display inflationary behaviour,
but the rate of inflation depends
on the value of $\phi _i$ \cite{r11,r28}. To find what is the most probable
value of $\phi _i$ given by the wave function, we have to use the prefactor
of the WKB solutions. This prefactor allows us to define a conserved current
given by the expression:
\begin{equation}
\label{116}
J^{\alpha }=|A|^2\nabla ^{\alpha }S
\end{equation}
Then, we can define the probability measure:
\begin{equation}
\label{117}
{\rm d}P=J^a{\rm d}\phi
\end{equation}
where $J^a$ is the component of the current associated with the coordinate
$a$ in the minisuperspace (for the complete discussion,
see e.g \cite{r11,r24,r29}). Inserting (\ref{109}), (\ref{110}) (\ref{111})
and (\ref{112}) in (\ref{117}), we obtain:
\begin{eqnarray}
\label{118}
{\rm d}P_{HH} &\propto & e^{\frac{8}{V}}{\rm d}\phi \\
\label{119}
{\rm d}P^{(CS)}_C &\propto & \frac{V^{-\frac{2}{3}}}{\cos ^2(\frac{4}{V}
-\frac{\pi }{12})}{\rm d}\phi \\
\label{120}
{\rm d}P^{(CS)}_V &\propto & {\rm d}\phi \\
\label{121}
{\rm d}P_V &\propto & e^{-\frac{8}{V}}{\rm d}\phi
\end{eqnarray}
The measure $P^{(CS)}_C$ diverges when $V_k=\frac{48}{\pi (12k+7)}$.
In what follows, we restrict our considerations to the values of the
scalar field such that $|V|<<1$ in order to be sure of the validity of the
semiclassical approximation (if $V(\phi )=m^2\phi ^2$ then $\phi <<10^4$
in Planck units \cite{r11,r30}). But all the singularities of $P^{(CS)}_C$
are contained in the range $V<2.5$. Then, in spite of its nice form,
the choice leading to $\Psi ^{(CS)}_C$ must not be considered. We would
like to emphasize that this problem would occur for all the wave functions
with classical change of signature except for $\Psi ^{(CS)}_V$: this is a
direct consequence of the Hamiltonian quantization with the change of
signature. Indeed, the denominator in the expression (\ref{97}) leads, when
we take its asymptotic form, to trigonometric functions which possess
lot of zeros and therefore produces divergencies. The probability that
$\phi _i$ must be greater than $\phi _{suf}$ (the
value of the scalar field sufficient to produce inflation in agreement
with present observations, $\phi _{suf}\approx 4.4$ \cite{r11,r30})
knowing that $\phi _i$ is smaller than $10^4$, is given by:
\begin{equation}
\label{122}
P^{(CS)}_V(\phi _i>\phi _{suf}|0<\phi _i<10^4)=
\frac{\int ^{10^4}_{\phi _{suf}}{\rm d}\phi }{\int ^{10^4}_0{\rm d}\phi }
\approx 1
\end{equation}
Sufficient inflation is predicted indeed by $\Psi ^{(CS)}_V$. It is easy to
show that sufficient inflation is also predicted by $\Psi _V$ but not by
$\Psi _{HH}$ \cite{r11}. We have seen that if we try to quantize General
Relativity with the classical change of signature, a unique boundary
condition that is acceptable is the tunneling boundary condition (contrary
to usual quantum cosmology where many choices are possible). The wave
function satisfying this condition predicts sufficient inflation. However, its
first derivative is not continuous when the change of signature occurs.

\section{Conclusion}

The aim of this article was to construct a modified version of the
Hamiltonian formulation of General Relativity in order to incorporate
classical solutions of Einstein's equations displaying the change of
signature. We have shown that Euclidean solutions correspond to a region in
the superspace with a signature of the supermetric being ($+-----$) instead
of ($-+++++$). The case of the minisuperspace describing the Robertson-Walker
cosmological solution with a scalar field has been studied in particular.
Next, we have quantized the theory in this minisuperspace. This has allowed
us to compute the wave function corresponding to classical Euclidean
solutions. We have also considered different boundary conditions. Only the
boundary condition introduced by Vilenkin leads to an acceptable behaviour
of the wave function. However, the derivative of the wave function is not
continuous when the change of signature occurs.

\section{Acknowledgments}

I would like to express my gratitude to L. P. Grishchuk for several
enlightning discussions concerning this work. It is also a pleasure to thank
R. Kerner for many useful discussions, constant encouragement and reading of
the text. I would also like to thank N. Pinto-Neto for introducing me to the
Dirac formalism and for enlightening discussions and remarks.

\appendix
\section{Appendix}

\setcounter{equation}{0}
In this appendix, we remind the basic formulae of the Hamiltonian formalism
of General Relativity in order to be able to compare them with the relations
including the change of signature considered at the first section of this
article. The metric is decomposed according to the expression:
\begin{equation}
\label{123}
{\rm d}s^2=(N^iN_i-N^2){\rm d}t^2+2N_i{\rm d}x^i{\rm d}t+h_{ij}{\rm d}x^i
{\rm d}x^j
\end{equation}
and the extrinsic curvature can be written as:
\begin{equation}
\label{124}
K_{ij}=-\frac{1}{2N}(^{(3)}\nabla _{(i}N_{j)}-\partial _t h_{ij})
\end{equation}
Then, using the previous equations, the computation of the Christoffel
symbols yields \cite{r31}:
\begin{eqnarray}
\label{125}
\Gamma^0_{00} &=& \frac{\stackrel{\mbox{\LARGE .}}{N}}{N}+\frac{N^iN_{,i}}{N}
+ \frac{N^iN^j}{N}K_{ij} \\
\label{126}
\Gamma^0_{0i} &=& \frac{N_{,i}}{N}+\frac{N^k}{N}K_{ik} \\
\label{127}
\Gamma^0_{ij} &=& \frac{1}{N}K_{ij} \\
\label{128}
\Gamma^i_{00} &=& -\frac{1}{2}h^{ik}(N_jN^j-N^2)_{,k}+Nh^{ik}\partial _{t}
              (\frac{N_k}{N})-\frac{N^iN^k}{N} N_{,k} \nonumber \\
              & &-\frac{N^iN^kN^m}{N}K_{km} \\
\label{129}
\Gamma^i_{j0} &=& N(K^i{}_j+^{(3)}\nabla(\frac{N^i}{N})-\frac{N^iN^k}{N^2}
K_{jk}) \\
\label{130}
\Gamma^k_{ij} &=& ^{(3)}\Gamma^k_{ij}-\frac{N^k}{N}K_{ij}
\end{eqnarray}
The components of the Ricci tensor are given by \cite{r19}:
\begin{eqnarray}
\label{131}
R_{00} &=& -Nh^{ki}\stackrel{\mbox{\LARGE .}}{K}_{ki}+{}^{(3)}
       \nabla _k(\partial ^kN)+2NK^i{}_{j}{}^{(3)}\nabla _iN^j \nonumber \\
       & & +2NN^{j}{}^{(3)}\nabla _k K^k{}_j-NN^j{}^{(3)}\nabla _j K^k{}_k
       +N^2K_{kj}K^{kj}
       \nonumber \\
       & & +N^lN^j{}^{(3)}R_{lj}+N^lN^jK_{lj}K^k{}_k-2N^iN^kK_{kj}K^j{}_i
       \nonumber \\
       & & +\frac{N^iN^k}{N}\stackrel{\mbox{\LARGE .}}{K}_{ki}-
       \frac{N^iN^k}{N}{}^{(3)}\nabla _k
       (\partial _i N)-\frac{2}{N}N^iN^kK_{kj}{}^{(3)}\nabla _iN^j
       \nonumber \\
       & & -\frac{1}{N}N^iN^kN^j{}^{(3)}\nabla _k K_{ij}
\end{eqnarray}

\begin{eqnarray}
\label{132}
R_{0i} &=& \frac{N^k}{N}\stackrel{\mbox{\LARGE .}}{K}_{ki}+N^j{}^{(3)}R_{ij}
       -\frac{N^k}{N}{}^{(3)}\nabla _k(\partial _i N)-N\partial _iK^j{}_j
       \nonumber \\
       & &+N{}^{(3)}\nabla _j(K^j{}_i)+N^kK_{ik}K^j{}_j
       -\frac{N^jN^m}{N}{}^{(3)}\nabla _jK_{im}
       \nonumber \\
       & &-\frac{N^k}{N}K_{jk}{}^{(3)}\nabla _iN^j
       -\frac{N^j}{N}K_{im}{}^{(3)}\nabla _jN^m-2N^jK_i{}^mK_{mj}
\end{eqnarray}

\begin{eqnarray}
\label{133}
R_{ij} &=& {}^{(3)}R_{ij}+\frac{1}{N}\stackrel{\mbox{\LARGE .}}{K}_{ij}-
       2K_{ki}K^k{}j+K^k{}_kK_{ij}
       \nonumber \\
       & & - \frac{N^k}{N}{}^{(3)}\nabla _kK_{ij}
       -\frac{1}{N}{}^{(3)}\nabla _j(\partial _iN)
       -\frac{1}{N}K_{ki}{}^{(3)}\nabla _j N^k
       \nonumber \\
       & &-\frac{1}{N}K_{jk}{}^{(3)}\nabla _iN^k
\end{eqnarray}
The Ricci scalar deduced from the above equations takes on the form:
\begin{eqnarray}
\label{134}
R &=& {}^{(3)}R-3K_{ki}K^{ki}+K^2-\frac{2}{N}N^i\partial _iK^j{}j
  \nonumber \\
  & & +\frac{2}{N}h^{ki}\stackrel{\mbox{\LARGE .}}{K}_{ki}
  -\frac{2}{N}{}^{(3)}\nabla _k(\partial ^kN)
  -\frac{4}{N}K_k{}^j{}^{(3)}\nabla _jN^k
\end{eqnarray}
One can also write $R$ as:
\begin{eqnarray}
\label{135}
R &=& {}^{(3)}R+K_{ki}K^{ki}+K^2+\frac{2\stackrel{\mbox{\LARGE .}}{K}}{N}
  -\frac{2N^i}{N}N^i\partial _i K
  \nonumber \\
  & &-\frac{2}{N}{}^{(3)}\nabla _k(\partial ^k N)
\end{eqnarray}
Finally, we obtain the well-known formula giving the Lagrangian of the
gravitational field including the surface term:
\begin{eqnarray}
\label{136}
{\cal L}_G &=& h^{\frac{1}{2}}N({}^{(3)}R+K_{ki}K^{ki}-K^2) \nonumber \\
           & & +2\frac{{\rm d}}{{\rm d}t}(h^{\frac{1}{2}}K)
           -2\partial _i(h^{\frac{1}{2}}KN^i+h^{\frac{1}{2}}h^{ki}
           \partial _kN)
\end{eqnarray}

\newpage
Caption of the figure:
\par
$\Psi _{HH}$, $\Psi _V$, $\Psi ^{(CS)}_C$ and $\Psi ^{(CS)}_C$ in the
space of Wave functions.

\end{document}